\documentclass[aps,prb,twocolumn,showpacs,preprintnumbers,amsmath,amssymb,superscriptaddress]{revtex4}
\usepackage{graphicx}% Include figure files
\usepackage{dcolumn}% Align table columns on decimal point
\usepackage{bm}% bold math

\begin{document}

\preprint{  }

\title{ Anomalous magnetoresistance of EuB$_{5.99}$C$_{0.01}$: Enhancement of magnetoresistance\\ in systems with magnetic polarons  }

%\title{Electric and magnetic transport properties of carbon doped EuB$_6$}
% Force line breaks with \\

\author{M. Bat$\!$'kov\'a }   
\email{batkova@saske.sk}  

	\affiliation{Institute  of  Experimental  Physics, Slovak  Academy 
 			 of Sciences, Watsonova 47, 04001~Ko\v {s}ice, Slovakia}
 			 \author{I. Bat$\!$'ko}%
\affiliation{Institute  of  Experimental  Physics, Slovak  Academy 
 			 of Sciences, Watsonova 47, 04001~Ko\v {s}ice, Slovakia} 			 
\author{K. Flachbart}%
	\affiliation{Institute  of  Experimental  Physics, Slovak  Academy 
 			 of Sciences, Watsonova 47, 04001~Ko\v {s}ice, Slovakia}
\author{K. Jurek}
			\affiliation{Institute of Physics, Academy of Sciences of the Czech Republic, Na Slovance 2, 182 21 Prague, Czech Republic }
\author{E. S. Konovalova}
	\affiliation{Institute for Problems of Material Science, NASU, 252680~Kiev, 
			Ukraine }
\author{J.~Kov\'a\v c}%
	\affiliation{Institute  of  Experimental  Physics, Slovak  Academy 
 			 of Sciences, Watsonova 47, 04001~Ko\v {s}ice, Slovakia}
\			
\author{M. Reiffers}%
 	\affiliation{Institute  of  Experimental  Physics, Slovak  Academy 
 			 of Sciences, Watsonova 47, 04001~Ko\v {s}ice, Slovakia}
\author{V. Sechovsk\'y}
	\affiliation{Faculty of Mathematics and Physics, Charles University, Ke~Karlovu~5, 121 16 Praha 2, Czech Republic}
\author{N. Shitsevalova}
	\affiliation{Institute for Problems of Material Science, NASU, 252680~Kiev, 
			Ukraine }	
\author{E. \v {S}antav\'a}
\author{J. \v {S}ebek}
	\affiliation{Institute of Physics, Academy of Sciences of the Czech Republic, Na Slovance 2, 182 21 Prague, Czech Republic }

\date{\today}% It is always \today, today,

\begin{abstract}     
We present results of measurements of electrical, magnetic and thermal 
	properties of EuB$_{5.99}$C$_{0.01}$.
The observed anomalously large negative  magnetoresistance as above, so below 
	the Curie temperature of ferromagnetic ordering $T_C$ is attributed to
	 fluctuations in carbon concentration.
Below $T_C$  the carbon richer regions give rise to helimagnetic domains, which are  
	responsible for an additional scattering term in the resistivity, 
	which can be suppressed by a magnetic field.
Above $T_C$ these regions prevent the process of  percolation of magnetic polarons (MPs), 
	acting as ``spacers" between MPs.
We propose that such ``spacers", being in fact volumes incompatible with  existence of MPs, 
	may be responsible for the decrease of the percolation temperature and
	for the additional (magneto)resistivity increase in systems with MPs.	
\end{abstract}
\pacs{75.30.Kz, 72.15.Gd, 75.47.Gk }
                           
\maketitle
EuB$_6$ is a rare example of a low carrier density hexaboride 
	that orders ferromagnetically  at low temperatures and 
	undergoes a metal-insulator phase transition. 
The ferromagnetic order is established via two consecutive
	phase transitions \cite{Degiorgi97,Sullow98,Sullow00a} 
	at $T_M$~=~15.5~K  and $T_C$~=~12.6~K,  respectively. \cite{Sullow00a}  
The high-temperature magnetoresistance
	 is large and negative; the absolute value is increasing  
	with decreasing temperature, and reaches the maximum value of about 100\%  
	at the magnetic ordering  temperature (15.6~K). \cite{Paschen00}  
In the ferromagnetic regime, the magnetoresistance is positive
	and reaches a value up to 700\% at 7~T and 1.7~K. \cite{Paschen00}
			
Physical properties of EuB$_6$ are thought to be governed by magnetic polarons 
	(MPs), which are in fact carriers localized in ferromagnetic clusters 
	embedded in a paramagnetic matrix.\cite{Sullow00a,Snow01,Calderon04,Yu06}
As suggested by S\"ullow et~al., \cite{Sullow00a} the magnetic phase transition at $T_M$ represents the emergence of 
	the spontaneous magnetization accompanied by metalization. 
At this temperature the MPs begin to overlap and form a conducting, 
	ferromagnetically ordered 
	phase that acts as a percolating, low-resistance path across the otherwise
	poorly conducting  sample.\cite{Sullow00a} 
With decreasing temperature, the volume fraction of the conducting 
	ferromagnetic phase expands, until the sample becomes a homogeneous 
	conducting bulk ferromagnet at $T_C$. \cite{Sullow00a} 
As indicated by Raman scattering measurements,\cite{Snow01} the
 	polarons appear in EuB$_6$ at about 30~K.

Because of the very low number of intrinsic charge carriers
	($\sim$10$^{20}$~cm$^{-3}$), \cite{Aronson99}
	even a slight change of the concentration of conduction electrons 
	(e.g. due to a change of chemical composition or	number of impurities)
	 can drastically modify the electric 
	and magnetic properties of EuB$_6$.\cite{Kasaya78,Molnar81}  
Substitution of B by C enhances the  charge carrier concentration in EuB$_6$.	
As shown by neutron diffraction studies,\cite{Tarascon81} the predominant
	ferromagnetic ordering in the stoichiometric EuB$_6$ changes with increasing 
	carbon content through a mixture of the ferromagnetic phase and helimagnetic
	domains into a purely antiferromagnetic state. 
 The paramagnetic Curie temperature
 	$\theta_p$ of EuB$_{6-x}$C$_{x}$ changes its sign for $x$~=~0.125. \cite{Kasaya78}
The helimagnetic domains are associated with carbon 
	richer regions (with  higher carrier density) due to
	 local fluctuations of the carbon concentration.
Different impact of the RKKY interaction in  carbon richer and carbon 
	poorer regions  yields to different  types of magnetic order. \cite{Tarascon81}
 
The unusual transport properties of carbon doped EuB$_6$ single crystal
	were reported  more than a decade ago.\cite{Batko95}  	
The results have shown that the electrical resistivity becomes strongly enhanced below 15~K and exhibits a maximum 	around 5~K.  
The 	residual resistivity is exceptionally high; it is even higher than the room temperature resistivity $\rho$(300~K).
Application of a magnetic field of 3~T at 4.2~K causes a dramatic reduction 
 	of the resistivity yielding $\rho$(0~T)/$\rho$(3~T)~=~3,7. 
The huge residual resistivity has been ascribed to the scattering of conduction electrons on
	boundaries between the ferromagnetic and helimagnetic regions.\cite{Batko95}

In this paper we present an extended study 
 	of the electrical resistivity, magnetoresistance,
 	 susceptibility and heat capacity on a EuB$_{5.99}$C$_{0.01}$ single crystal.
We bring further experimental results supporting the afore mentioned hypothesis of 	
	the dominant scattering process at temperatures below $T_C$ 
	originating from the mixed magnetic structure.
In addition, our results advert that above $T_C$
	the electrical transport is governed by MPs and can be well understood within a recently proposed 
	scenario involving the ``isolated'', ``linked'' and ``merged'' MPs.\cite{Yu06} 
Moreover, we argue that regions of proper size and space distribution,  
	 incompatible with existence of MPs can be the clue for understanding
	 the origin of the colossal magnetoresistance in systems with MPs.       
				
Samples used for magnetization and resistivity measurements  
	were cut from the crystal used in previous studies, \cite{Batko95}
which has been
	grown by means of the zone-floating. 
Recent micro-probe analysis of this crystal revealed the carbon content corresponding 
	to the stoichiometric formula EuB$_{5.99}$C$_{0.01}$.
The electrical resistance, magnetoresistance, heat capacity and ac-susceptibility were measured
	 in the Quantum Design PPMS and MPMS. 
The direction of the applied magnetic field was perpendicular to 
	electrical current in all magnetoresistance measurements. 	
	\begin{figure}[!t]%
	\center{
	\resizebox{0.96\columnwidth}{!}{%
	\includegraphics{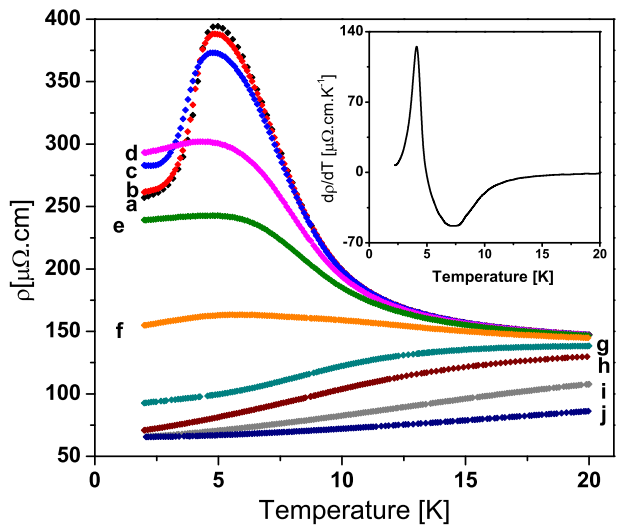}}}
	 \caption{Temperature dependences of the 
	resistivity of EuB$_{5.99}$C$_{0.01}$ in magnetic fields 
	a) 0~T, b) 50~mT, c) 0.1~T, d) 0.3~T, e) 0.5~T, 
	f) 1~T, g) 2~T, h) 3~T, i) 6~T, and j) 12~T. 
	The inset shows the resistivity derivative  $d\rho/dT$ at 0~T. } 			\label{RTB}
			\end{figure}	
															
The electrical resistivity of EuB$_{5.99}$C$_{0.01}$ decreases upon
		cooling from 300~K until it reaches a shallow minimum at about 40~K.
Below 10~K it increases steeply, passes a maximum at $T_{RM}$~$\sim$5~K, and subsequently
	falls off having tendency to saturate at lowest temperatures.
The low-temperature part of its dependence is shown
	in Fig.~\ref{RTB} as curve a).
	The temperature derivative of the resistivity  in zero magnetic field, 
	depicted in the  figure inset, shows a sharp maximum 	at $T_m$~=~4.1~K
	indicating a  proximity of  magnetic phase-transition.
Since the optical reflectivity data of the studied 
    system have not revealed any shift in the plasma frequency  
    between 4.2 and 20~K,\cite{Batko95} the charge carrier concentration
    can be regarded as constant in this temperature interval.
Therefore, we tentatively associate the anomalous 
    resistivity behavior with magnetism in this material.		
	\begin{figure}[!t]%[p]%
	\center{
	\resizebox{0.97\columnwidth}{!}{%
	\includegraphics{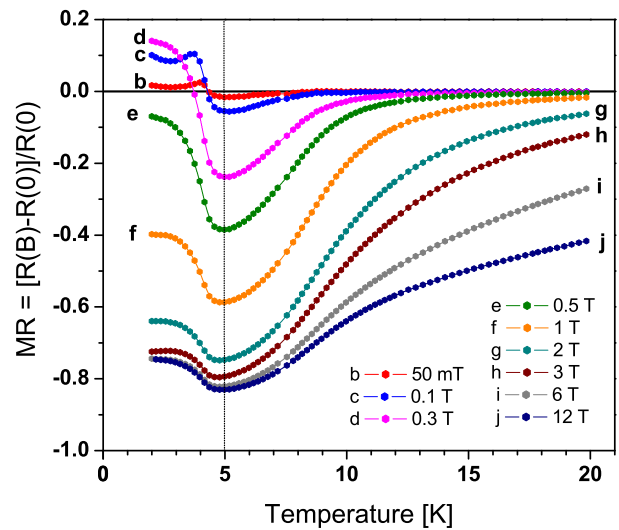}
		}}
	\caption{Temperature dependences of  the
	magnetoresistance $MR = [\rho(B)-\rho(0)]/\rho(0)$
	derived from the data in Fig. \ref{RTB}. %The solid lines are to guide the eye.
	}
	\label{MRT}
	\end{figure}

In Fig.~\ref{MRT} we plot temperature dependences of magnetoresistance
	$MR=[\rho(B)-\rho(0)]/\rho(0)$ for selected values of the applied magnetic 
	field between  50~mT  and 12~T, derived from the data shown in Fig.~\ref{RTB}.
The absolute value of magnetoresistance reaches a maximum (of about 0.83 for 12~T) in the vicinity of $T_{RM}$.
Upon further cooling $MR$ decreases continuously in absolute value,
     and in difference to EuB$_6$, only in the smallest magnetic fields up to 0.3~T,
    and at the lowest temperatures, it passes through zero and reaches positive 
    values. 

We suppose that below $T_C$, the scattering of conduction electrons originates
	from phase boundaries of the mixed magnetic structure consisting of 
	helimagnetic domains, associated with carbon richer regions, in the ferromagnetic matrix.
Sufficiently high magnetic field makes the
	helimagnetic domains energetically unfavorable and therefore reduces 
	their volumes (and probably destroys them completely at highest fields), giving rise to 
	negative magnetoresistance.

\begin{figure}[!t]%
	\center{
	\resizebox{0.99\columnwidth}{!}{%
	\includegraphics{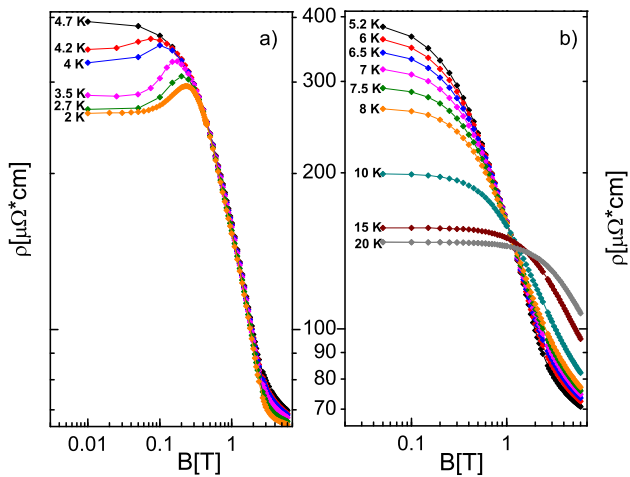}
		}}
\caption{Resistivity dependence on the magnetic field measured at selected 
	temperatures below (a) and above (b) the temperature of the resistivity maximum.} 	\label{RBT}
	\end{figure}

The magnetic field influence on the resistivity and magnetoresistance behavior between 
	2 and 20~K and the magnetic field dependences of resistivity
	depicted in Fig.~\ref{RTB}, \ref{MRT} and \ref{RBT} respectively, reveal
	two different magnetoresistance regimes:	
(i)	for temperatures lower than  $T_{RM}$ -
	the resistivity is enhanced by small fields ($B$~$\leq$ 0.3~T) and 
	 reduced by higher fields ($B$~$\geq$ 0.5~T); 
(ii) above $T_{RM}$ - the resistivity monotonically decreases with increasing 
applied 
magnetic field.

\begin{figure}[!b]
	\center{
	\resizebox{0.96\columnwidth}{!}{%								
	\includegraphics{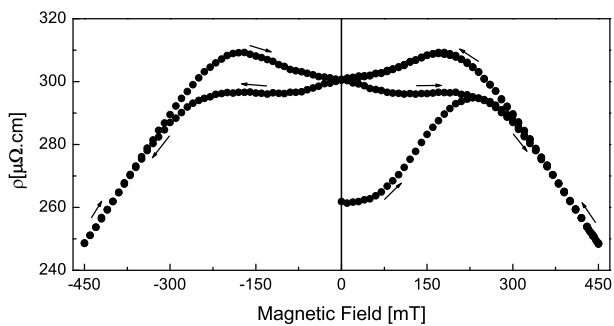}
	 }}
\caption{Hysteresis of magnetoresistance of EuB$_{5.99}$C$_{0.01}$ at 2~K. 
After cooling at in zero magnetic field, the field was increased up to 450~mT, 
	then decreased to -450~mT, and finally increased to 450~mT.
	 The arrows 	show the emergence of the curve.}
	\label{Hyst}
	\end{figure}

The low-field magnetoresistance measured at 2~K is dependent on magnetic history and exhibits 
	large hysteresis.
Fig.~\ref{Hyst} shows the hysteresis behavior 
	   of  the resistivity, including the virgin  
	   curve taken at 2~K after cooling from 30~K to 2 K in zero magnetic field.
As it is visible in the figure, the hysteresis is significant 
	for $\left|B\right|$$\leq$~0.3~T.
The hysteresis of magnetisation is very weak, but not negligible 
	in the interval where the resistivity hysteresis is observed, 
	suggesting that the positive magnetoresistance in low magnetic fields is due to the  conduction-electron scattering on the domain walls within the ferromagnetic matrix.

With the aim to get more information on the magnetic properties and the phase transition(s), 
	we measured the real part of the temperature dependence of the ac-susceptibility 
$\chi^,(T)$ and the specific heat $C(T)$ in the temperature range 2~-~86~K and 
	2~-~30~K, respectively.   
The $1/\chi^,(T)$ satisfies the Curie-Weiss law in 
	the region above $\sim$29~K and yields the paramagnetic Curie temperature
	$\theta_p$~=~7~K.
Fig.~\ref{Susc} shows the $\chi^,(T)$  and $C(T)$ data below 10~K.
The $\chi^,(T)$ dependence indicates two distinct regimes, one above and other below $\sim 4~K$,
	 such as it obeys almost linear behavior, in the intervals 2~-~3.6~K and 4.1~-~4.8~K, 
	 respectively, however with different slopes.
The  specific heat exhibits a broad peak at 5.7~K, which we tentatively associate with 	 the magnetic ordering transition at $T_C$ $\sim$ 5.7 K.
The position of the peak correlates well with the position of the inflexion 
	point of the $\chi^,(T)$ dependence (5.5~K).
There is also a side anomaly at 4.3~K in the $C(T)$ dependence, which almost coincides with the  
	afore mentioned resistivity anomaly at $T_m$~=~4.1~K and with the	change of the regime of the $\chi^,(T)$ 		dependence.
Detailed microscopic investigation (e.g. neutron diffraction) is desired to 		elucidate the relation 
	 between the specific-heat and resistivity anomalies with magnetic phenomena 
	 in the studied material.
	 
	\begin{figure}[!t]
			\center{
			\resizebox{0.93\columnwidth}{!}{%						
			\includegraphics{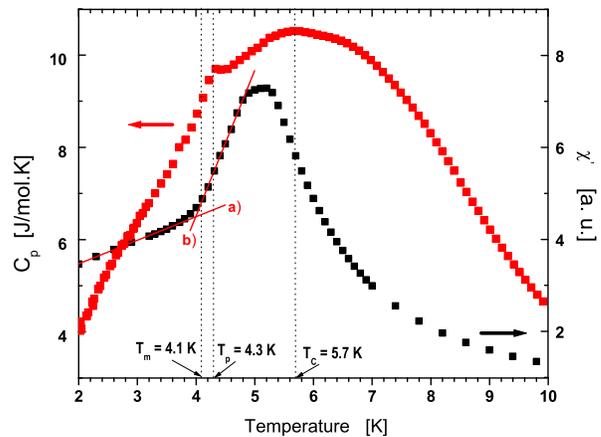}
					}}
\caption{Temperature dependence of the real part of the ac-susceptibility and the specific heat.
The red lines a) and b) are linear fits for the temperature intervals  2~-~3.6~K and 4.1~-~4.8~K,  respectively.}
			\label{ACS}
			\end{figure}

The observed behavior of EuB$_{5.99}$C$_{0.01}$
 	can be consistently explained within the framework of results obtained 
 	by Yu and Min \cite{Yu06}, who investigated the
   	magnetic phase transitions in MP systems using the Monte Carlo method.
They supposed three consecutive temperature scales: $T^{*}$, $T_{C}$ and 
   	$T_{p}$.
Upon cooling from the high-temperature paramagnetic state 
 	the isolated MPs with random magnetization directions  
 	 begin to form at $T^{*}$. \cite{Yu06} 
At further cooling the MPs grow in size.
Down to  $T_C$ carriers 
	are still confined to 
	MPs, thus the metallic and magnetic regions are 
	separated from the insulating and paramagnetic regions. \cite{Yu06}
The isolated MPs become linked at the bulk ferromagnetic transition temperature $T_C$.
	Eventually, the polaron percolation occurs expressing itself as 
	a peak in the heat capacity at $T_p$~$<$~$T_C$. \cite{Yu06}
Below $T_p$ all carriers are fully delocalized and the concept of MPs becomes
	 meaningless.
The other issue, which should be mentioned, is that the impurities
	reduce both, $T_C$ and $T_p$, but the discrepancy ratio ($T_p/T_C$ = 7/9 $\doteq$ 0.7$\overline{7}$)
	 between these two temperatures is retained. \cite{Yu06}

According to the concept of Yu and Min,\cite{Yu06} we interpret the 
	obtained experimental results as follows. 
Consistently with the temperature dependence
	of the magnetization, EuB$_{5.99}$C$_{0.01}$ is paramagnetic above 
	$\sim$29~K. 
We expect the formation of isolated MPs at lower temperatures. 
The magnetic phase transition temperature reflected in the broad maximum of the $C(T)$ dependence, 
	we associate with the temperature of ferromagnetic ordering $T_{C}$.
We suggest that the isolated MPs begin to link at $T_{C}$. 
The MPs become merged and percolation occurs at the temperature of the (side) specific-heat anomaly $T_p$ = 4.3~K. 
Here is an excellent correspondence between the theoretically expected ratio 
	 $T_p$/$T_C$ $\doteq$ 0.7$\overline{7}$ and our experimental value 
	 $T_p$/$T_C$ = 4.3/5.6 $\doteq$~0.75. 
The transition to the percolated phase is accompanied by the abrupt increase of the electrical
	 conductivity of the percolated/merged phase at $T_m$~=~4.1~K. 
The fact that  $T_m$ is lower than both $T_p$ and $T_C$
	 supports the supposition that Fisher-Langer relation \cite{Fisher68}
	is not valid in MP systems because of spatial inhomogeneity. \cite{Yu06}

The concept outlined above allows us also to explain the very interesting 
	issue connected with the larger value of the resistivity maximum observed
	for EuB$_{5.99}$C$_{0.01}$ ($\sim$390~$\mu\Omega$~cm at $\sim$5~K) 
	than for EuB$_6$ ($\sim$350~$\mu\Omega$~cm at $\sim$15~K), \cite{Sullow98}
	despite  EuB$_{5.99}$C$_{0.01}$ is at room temperature 
	about four times better conductor,
 	having $\rho$(300~K)$\sim$180~$\mu\Omega$.cm,
	than EuB$_6$ with	$\rho$(300~K)$\sim$730~$\mu\Omega$.cm.\cite{Sullow98}
Since MPs can exist only in low carrier density environment, we suggest that 
	the carbon richer regions with an enhanced carrier concentration act as ``spacers'' 
	between MPs preventing them to link and merge. 
As a consequence, the system persists in poorly conducting state down to
	lower temperatures. 
Due to the extension of the temperature interval, in which the resistivity increases
	with decreasing temperature,
  	an additional resistivity increase is observed, resulting 
  	in the higher value of the resistivity maximum (and the larger negative magnetoresistance).
It may be generalized that the processes preventing MPs to link and to 
 	percolate extend the region of thermally activated 
 	transport (governed by MPs) to lower temperatures, giving rise to a higher
 	value of the resistivity maximum, resulting in the higher magnetoresistance. 

It seems that the colossal magnetoresistance of Eu$_{1-x}$Ca$_x$B$_6$  \footnote{see e.g. \cite{Wigger04-PRL,Wigger02,Paschen00}} might be also explained by this scenario
	assuming that the calcium richer regions play  a role similar to the
	carbon richer regions in EuB$_{6-x}$C$_x$.   
From the point of view of tuning the properties 
	of magnetoresistive materials here arises an interesting anology between
	the role of non-ferromagnetic ``spacers" in the  magnetoresistance enhancement in this class of materials,
	and the role of the (non-superconducting) pinning centers in the increase of the critical field in  superconductors.	

In summary, our studies reveal a large 
	negative magnetoresistence of  EuB$_{5.99}$C$_{0.01}$ well
	above and bellow the temperature of the bulk ferromagnetic ordering.
In the temperature region, where the resistivity maximum appears and transport 
	properties are governed by MPs, the results 
	have been consistently explained within the picture of isolated, 
	linked and merged MPs. \cite{Yu06}
We have observed three distinctive temperatures: $T_C$~=~5.7~K, 
	$T_p$~=~4.3~K and $T_m$~=~4.1~K.
As we suppose, at $T_C$ and $T_p$, being the temperatures of heat capacity maxima,
	 the isolated MPs become linked  and merged, respectively.
The peak in the $d\rho/dT$ vs. $T$ dependence at $T_m$, lying slightly below the percolation 
	temperature, we consider as a sign of rapid enhancement of the conductivity in the merged phase.
The unusually high value of the electrical resistivity  maximum 
	we associate with  the presence of carbon richer regions.
We suppose that these regions are responsible for the higher value of the resistivity maximum at 
	correspondingly lower temperatures, and consequently, for the larger magnetoresistance.
Finally, we emphasize  that introducing such ``spacers", which prevent the percolation of MPs 
may strongly enhance the magnetoresistance 
 	of systems with transport governed by MPs.
 	The ``spacers" are in fact regions of appropriate size and scale distribution, which are not compatible with ferromagnetic ordering.
This might show a route for future research efforts in relation 
	with the colossal magnetoresistance effect.

 This work was supported by 
the Slovak Scientific Agency VEGA (Grant No.~2/7184/27), the Slovak Research and Development Agency (Project No.~APVT-51-031704), and by the European Science Foundation (COST Action P16).
 The work of V. S. is a part of the research plan MSM 0021620834 that is financed
  by the Ministry of Education of the Czech Republic.

%\bibliographystyle{elsart-num}  
%\bibliography{references,acknow}

\begin{thebibliography}{10}
\expandafter\ifx\csname url\endcsname\relax
  \def\url#1{\texttt{#1}}\fi
\expandafter\ifx\csname urlprefix\endcsname\relax\def\urlprefix{URL }\fi

\bibitem{Degiorgi97}
L.~Degiorgi, E.~Felder, H.~R. Ott, J.~L. Sarrao, Z.~Fisk, Phys. Rev. Lett.
  79~(25) (1997) 5134.

\bibitem{Sullow98}
S.~S\"ullow, I.~Prasad, M.~C. Aronson, J.~L. Sarrao, Z.~Fisk, D.~Hristova,
  A.~H. Lacerda, M.~F. Hundley, A.~Vigliante, D.~Gibbs, Phys. Rev. B 57 (1998)
  5860.

\bibitem{Sullow00a}
S.~S\"ullow, I.~Prasad, M.~C. Aronson, S.~Bogdanovich, J.~L. Sarrao, Z.~Fisk,
  Phys. Rev. B 62~(17) (2000) 11626.

\bibitem{Paschen00}
S.~Paschen, D.~Pushin, M.~Schlatter, P.~Vonlanthen, H.~R. Ott, D.~P. Young,
  Z.~Fisk, Phys. Rev. B 61~(6) (2000) 4174.

\bibitem{Snow01}
C.~S. Snow, S.~L. Cooper, D.~P. Young, Z.~Fisk, A. Comment, J.~P. Ansermet, Phys. Rev. B 64 (2001) 174412.

\bibitem{Calderon04}
M.~J. Calder\'on, L.~G.~L. Wegener, P.~B. Littlewood, Phys. Rev. B 70 (2004)
  092408.

\bibitem{Yu06}
U.~Yu, B.~I. Min, Phys. Rev. B 74 (2006) 094413.

\bibitem{Aronson99}
M.~C. Aronson, J.~L. Sarrao, Z.~Fisk, M.~Whitton, B.~L. Brandt, Phys. Rev. B
  59~(7) (1999) 4720--4724.

\bibitem{Kasaya78}
M.~Kasaya, J.~M. Tarascon, J.~Etourneau, P.~Hagenmuller, Mat. Res. Bull. 13
  (1978) 751.

\bibitem{Molnar81}
S.~von Molnar, J.~M. Tarascon, J.~Etourneau, J. Appl. Phys. 52~(3) (1981) 2158.

\bibitem{Tarascon81}
J.~M. Tarascon, J.~L. Soubeyroux, J.~Etourneau, R.~Georges, J.~M.~D. Coey,
  O.~Massenet, Solid State Commun. 37 (1981) 133.

\bibitem{Batko95}
I.~Bat$\!$'ko, M.~Bat$\!$'kov\'a, K.~Flachbart, D.~Macko, E.~S. Konovalova,
  Y.~B. Paderno 140-144 (1995) 1177--1178.

\bibitem{Fisher68}
M.~E. Fisher, J.~S. Langer, Phys. Rev. Lett. 20 (1968) 665.

\bibitem{Wigger04-PRL}
G.~A. Wigger, C.~Beeli, E.~Felder, H.~R. Ott, A.~D. Bianchi, Z.~Fisk, Phys.
  Rev. Lett. 93~(14) (2004) 147203.

\bibitem{Wigger02}
G.~A. Wigger, C.~W\"alti, H.~R. Ott, A.~D. Bianchi, Z.~Fisk, Phys. Rev. B 66
  (2002) 212410.

\end{thebibliography}
\end{document}